\documentclass[a4paper,11pt]{article}
\pdfoutput=1 % if your are submitting a pdflatex (i.e. if you have
\usepackage{jinstpub} % for details on the use of the package, please
\usepackage{xcolor}
\usepackage{lineno}
\usepackage{hyperref}
\usepackage{graphicx}
\usepackage{subcaption}
\title{Detection of MeV electrons using a charge integrating hybrid pixel detector}
\author[a,1]{E. Fr\"ojdh\note{Corresponding author.}}\author[a]{F. Baruffaldi} \author[a]{A. Bergamaschi}  \author[a]{M. Carulla} \author[a]{R. Dinapoli} \author[a]{D. Greiffenberg} \author[a]{J. Heymes} \author[a]{V. Hinger} \author[a]{R. Ischebeck} \author[b]{S. Mathisen} \author[b]{J. McKenzie}   \author[a]{D. Mezza} \author[a]{K. Moustakas} \author[a]{A. Mozzanica}\author[a]{B. Schmitt}  \author[a]{J. Zhang}
\affiliation[a]{Paul Scherrer Institut, Forschungsstrasse 111, CH-5232 Villigen PSI, Switzerland}
\affiliation[b]{ASTeC, STFC Daresbury Laboratory, Warrington, WA4 4AD, UK}
\emailAdd{erik.frojdh@psi.ch}

\abstract{
Electrons are emerging as a strong complement to X-rays for diffraction based studies. In this paper we investigate the performance of a JUNGFRAU detector with 320 um thick silicon sensor at a pulsed electron source. Originally developed for X-ray detection at free electron lasers, JUNGFRAU features a dynamic range of 120 MeV/pixel (implemented with in-pixel gain switching) which translated to about 1200 incident electrons per pixel and frame in the MeV region. We preset basic characteristics such as energy deposited per incident particle, resulting cluster size and spatial resolution along with dynamic (intensity) range scans. Measurements were performed at 4, 10 and 20 MeV/c. We compare the measurements with GEANT4 based simulations and extrapolate the results to different sensor thicknesses using these simulations.
}


\proceeding{23$^{\text{nd}}$ International Workshop on Radiation Imaging Detectors\\
  June 26 - 30$^{\text{th}}$, 2022\\
  Riva del Garda, Italy}

\begin{document}
\maketitle
\flushbottom

\section{Electron diffraction with MeV electrons}

Electrons have been used for microscopy for 90 years, and the resolution quickly surpassed that of light microscopes. Today, transmission electron microscopes can achieve atomic resolution \cite{SpringerHandbook}. In addition, the use of electron beams for structure determination through diffraction has received considerable interest in recent years \cite{nanocrystallography}.
A coherent electron beam scatters on the sample, and the detector is placed in far field at the Fraunhofer plane. For crystalline samples, the structure of the molecule can be determined, making use of the methods developed for X-ray diffraction.

Typically, electrons with kinetic energies of a few hundred keV are used \cite{Naydenova2019, Clabbers2021}.  These non-relativistic electrons can be generated using DC sources, and continuous beams with excellent coherence can be obtained from field emitter sources. For certain applications however, there are benefits for the use of electron energies up to a few mega electron volt (MeV) \cite{MeVdiffraction} as these relativistic electrons offer a number of unique benefits:

\begin{itemize}
   \item Lower interaction probability results in greater penetration depth such that thicker samples can be studied.
   \item Relativistic effects partially compensate the space charge between the electrons such that more intense and shorter pulses can be generated.
   \item Pump probe experiments can increase their timing accuracy by utilizing, electrons with velocities close to the speed of light.
   \item The smaller de Broglie wavelength of MeV electrons allows for higher resolution.
\end{itemize}

\section{The JUNGFRAU detector}

Measuring at a pulsed source requires an integrating detector (as opposed to counting) since all particles arrive at the same time (${\sim}10^{-15}-10^{-12}s$). In this study we evaluate response of the JUNGFRAU \cite{Mozzanica2016} detector, which, is a charge integrating hybrid pixel detector with MeV electrons. It was originally developed for the SwissFEL free electron laser. With in-pixel dynamic gain switching it provides low noise (83~e$^-$ RMS in G0) and high dynamic range (120~MeV/pixel/frame). A readout chip consists of 256 x 256 pixels with 75 x 75~$\mu$m$^2$ size. Modules are then built from 2x4 chips bump bonded to a single 4 x 8~cm$^2$, 320 $\mu m$ thick silicon sensor. At time of publication the largest JUNGFRAU detector in use consists of 32 modules comprising 16~megapixels. More on JUNGFRAU can be found in \cite{Mozzanica2018}, and the in-pixel gain switching is described in \cite{Mozzanica2014, JungmannSmith2014}. In this publication, we aim to assess the capability of using JUNGFRAU for direct electron detection at a pulsed MeV source (\ref{sec:clara}), which could extend the application range beyond high-performance X-ray detection at free electron lasers and synchrotron sources. 
\section{Results}

\subsection{Single Electron Response}
\label{sec:cluster}
To characterize the basic performance of the JUNGFRAU detector and estimate the precision with which we can measure the number of incident electrons, we started with sparse ($\ll$ 1e-/pixel) illumination and looked at clusters (\ref{sec:cluster}) from individual electrons. This measurement posed a significant challenge since the initial beam is small (<mm$^2$) and there are many electrons in the bunch (>10 000). By using the scattering in the exit window (0.5 mm Be) and transporting beam through 1.22 m air the resulting intensity was low enough.

The average energy deposition per electron is measured to 120 keV at 4 MeV incident beam and the most probable value 84 keV. This gives a dynamic range of around 1000 primary electrons per frame (i.e. average energy deposition divided with the dynamic range), for a 320 $\mu m$ thick Silicon sensor. Simulations (\ref{sec:sim}) agree well with the measured values showing only a slight over representation for 2 pixel clusters.

\begin{table}[]
    \centering
    \begin{tabular}{ | c | c|c | c | c |}
    \hline
         Type & Thickness [$\mu m$] &Energy [MeV] & Edep [keV] & Size [pixels]\\
         \hline
         Meas & 320 & 4 & 120 & 2.46 \\
         Meas & 320 & 10 & 116 & 1.76 \\
         Meas & 320 & 20 & 112 & 1.56 \\
         \hline
         Sim & 320 & 4 & 115 & 2.30 \\
         Sim & 200 & 4 & 69 & 1.64 \\
         Sim & 100 & 4 & 33 & 1.18 \\
    \hline
    \end{tabular}
    \caption{Average energy deposition and cluster size from measurements and simulations.}
    \label{tab:par}
\end{table}

\begin{figure}
    \centering
    \includegraphics[width=\textwidth]{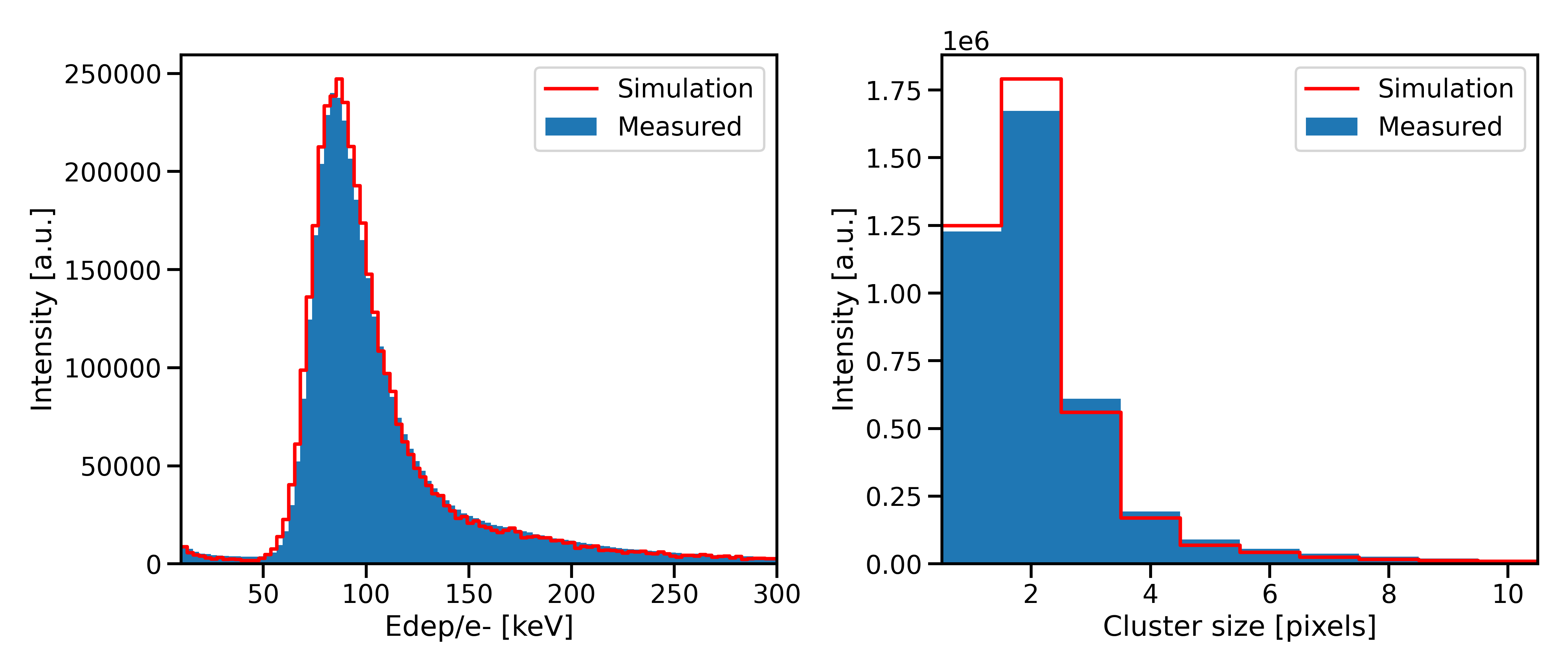}
    \caption{Simulations and measurements match overall, with only a slight over representation of two pixel clusters in the simulated data. The mean deposited charge per 4 MeV electron is 120 keV with the most probable value at 84 keV. (Error bars too small to be visible)}
    \label{fig:cluster}
\end{figure}

\begin{figure}
    \centering
    \includegraphics[width=\textwidth]{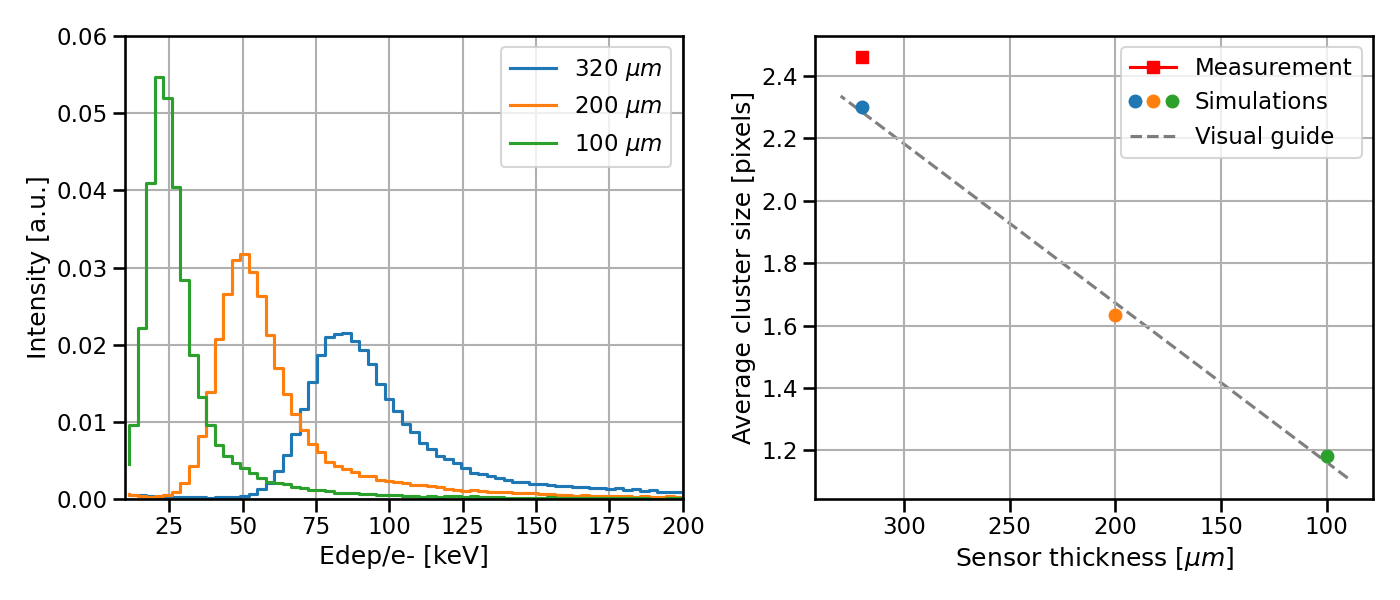}
    \caption{Simulated energy deposition and cluster size (4 MeV beam) as a function of sensor thickness. At 200 and 100 $\mu m$ the mean energy deposition per primary e- is 70 and 33 keV respectively.}
    \label{fig:sim_clust}
\end{figure}

Further, using simulations we can also investigate the effect of employing a thinner sensor. Figure \ref{fig:sim_clust} shows the simulated energy deposition and the average cluster size for 200 and 100 $\mu m$ thick sensors in addition to the 320 $\mu m$ sensor used in the experiments. Average energy deposition in the 200 and 100 $\mu m$  thick sensors is 70 and 33 keV respectively, which amounts to an increase of dynamic range to 1700 and 3600 primary e- (at 4MeV). With the thinner sensor we also see a reduction in cluster size, as expected, due to less scattering and lower diffusion. At 100 $\mu m$ the average cluster size is just 1.2 pixels, indicating a very good spatial resolution.  The simulated and measured mean energy depositions and cluster sizes are summarized in table \ref{tab:par}.

\subsection{Slanted Edge MTF measurements}

Cluster size gives an indication of the spatial resolution but to  quantify it we measured the Modular Transfer Function (MTF) using a slanted edge (\ref{sec:mtf}). At 4 MeV the resolution is degraded by scattering in the sensor layer while at 10 MeV we are already approaching the ideal MTF for a pixelated detector, which is given by $sinc(\pi \omega)$ \cite{Schroeder1999}. Looking at the simulations we observe that using a thinner sensor improves the spatial resolution due to less scattering in the sensor layer. 

The discrepancy between simulation and measurement can be attributed to the measurement setup. Using a thick edge it is always difficult to align in with the sensor. The beam was also not perfectly parallel since it originates as a point source on the Be window and is further scattered in air. These two factors combined with the penetration depth and scattering of 4 MeV electrons in the W edge leads to an under estimation of the spatial resolution.

\begin{figure}
    \centering
    \includegraphics[width=0.8\textwidth]{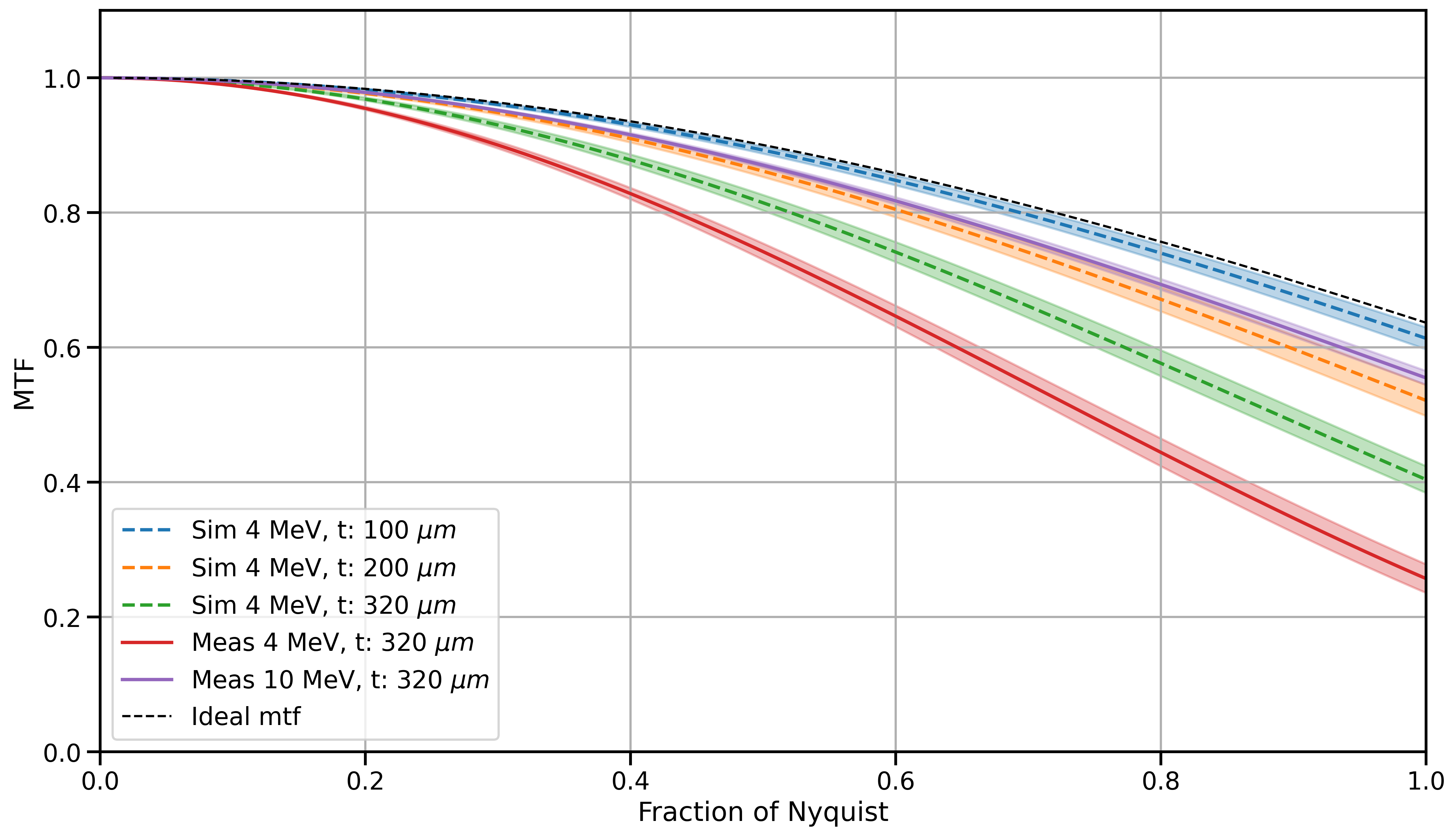}
    \caption{Simulated and measured MTF. For 4 MeV the spatial resolution is degraded by multiple scattering in the sensor layer, but either higher energies or thinner sensors improve the resolution. The difference between simulation and measurement can be attributed to the measurement setup.}
    \label{fig:mtf}
\end{figure}

\subsection{Charge Scans}

\begin{figure}
    \centering
    \includegraphics[width=0.8\textwidth]{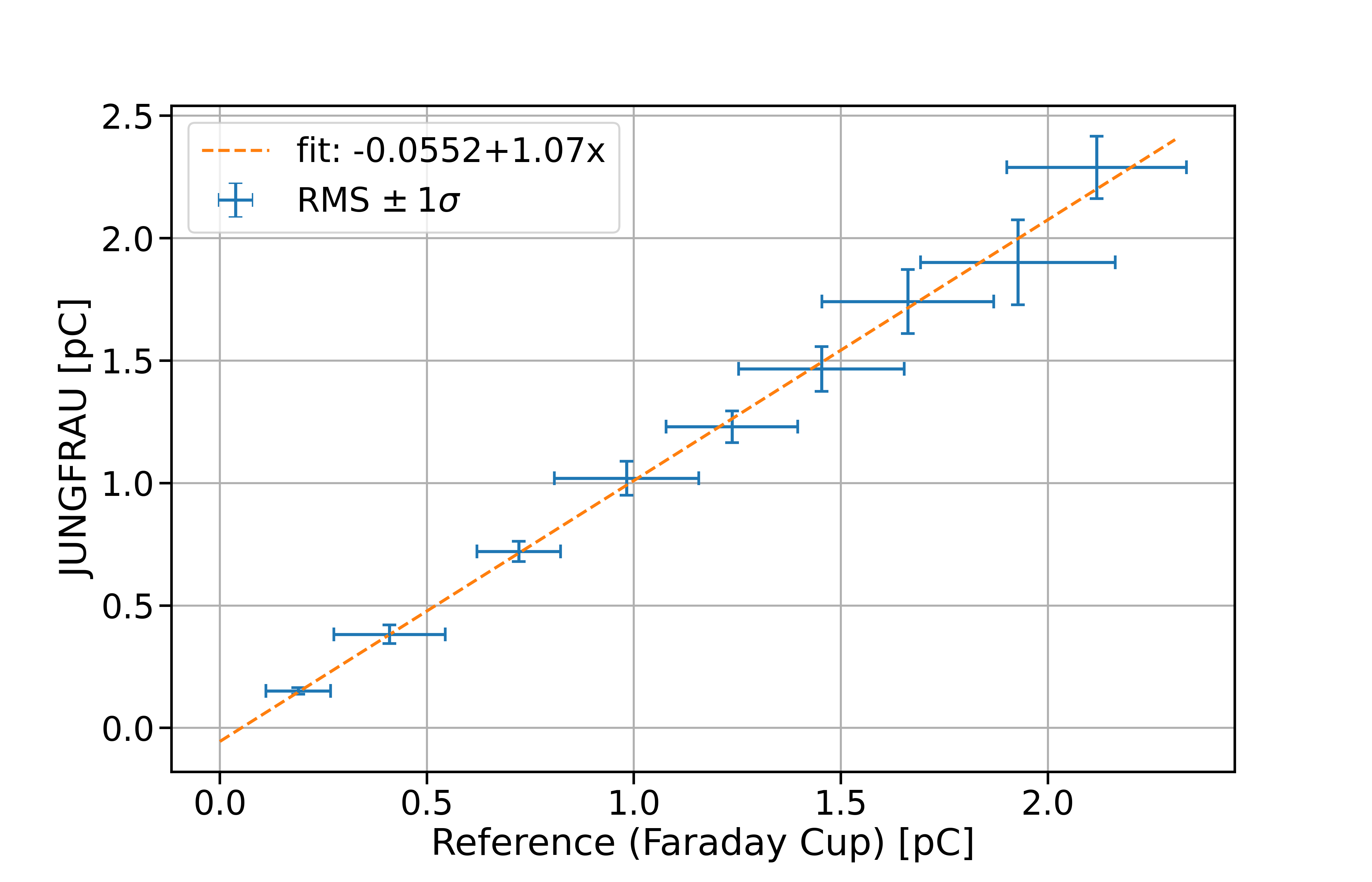}
    \caption{Charge measured with JUNGFRAU vs. reference detector. RMS value for 100 pulses with 1$\sigma$ error bars. In the last point we have around 100 MeV/pixel/frame in the beam. The small mismatch in charge could come from differences in sensitive area or calibration of either detector.}
    \label{fig:charge}
\end{figure}

To verify the linearity and calibration of the JUNGFRAU we swept the bunch charge over the full dynamic range. As a reference a custom built Faraday Cup was used. Figure \ref{fig:charge} shows the measured charge of the Faraday Cup compared with JUNGFRAU. For each step we used 100 pulses and plot the RMS value with 1$\sigma$ error bars. The uncertainty comes both from the shot to shot variation in the bunch charge as well as detector noise. For lower charges JUNGFRAU displays a much lower variation than the reference detector due to its lower noise but at higher charges both detector approach the same value since they are now presumably dominated by shot to shot variations.

\begin{figure}
    \centering
    \includegraphics[width=0.8\textwidth]{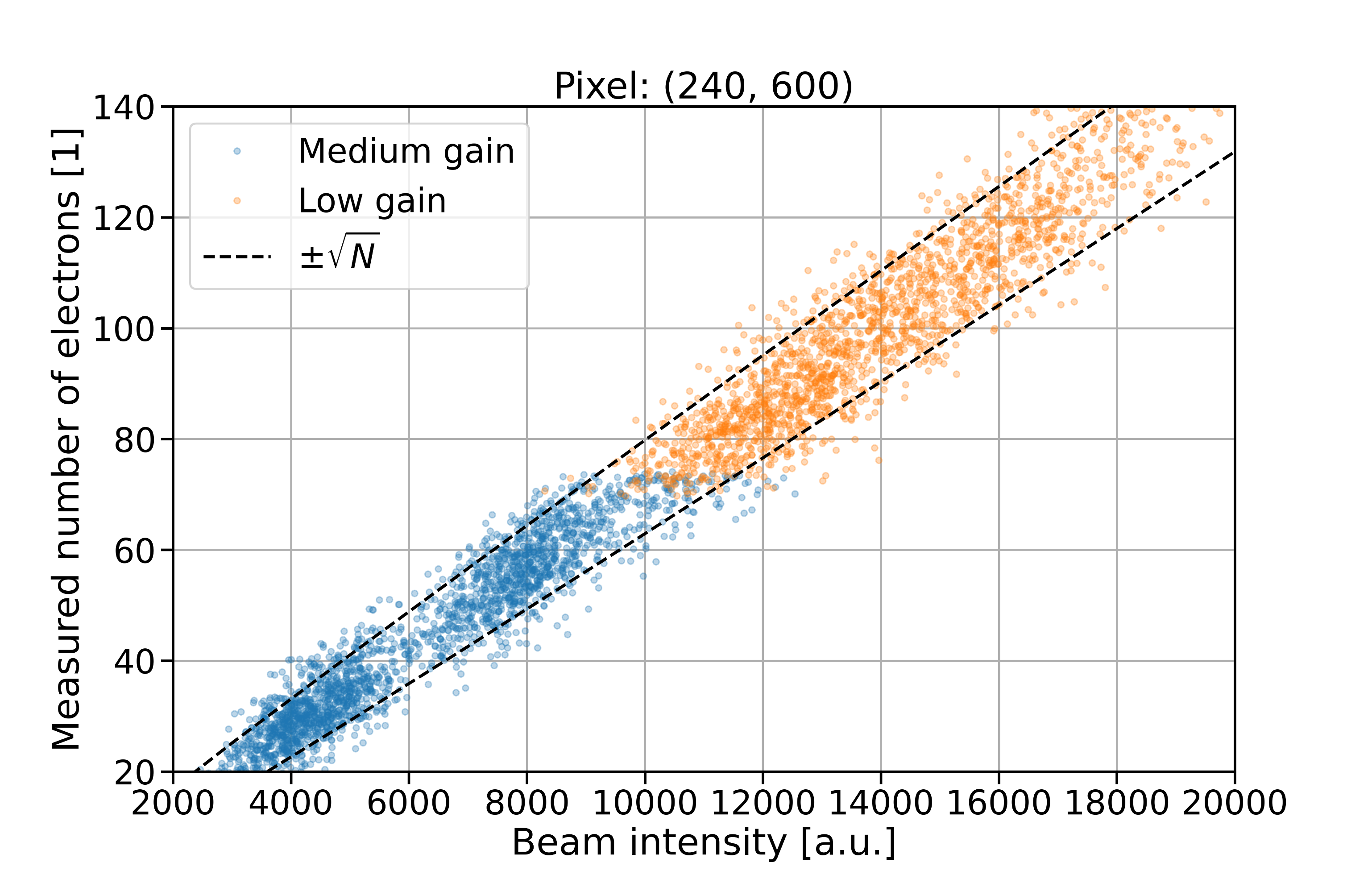}
    \caption{Charge scan over the gain switching range (medium->low) for pixel (240,600). Normalized using a low signal region of JUNGFRAU. The dashed guides show the expected variation due to the Poisson distribution of the incoming particles}
    \label{fig:pixel_scan}
\end{figure}

With a gain switching detector it is important that there are no artifacts or gaps in the gain switching region. Figure \ref{fig:pixel_scan} shows a single pixel switching between medium and low gain. As reference a low intensity (not switching) region of the JUNGFRAU was used. Ideally an independent detector should have been used, but no shot to shot measurements with sufficient accuracy were available. However, taking into account the single electron response measured in \ref{sec:cluster}, results from figure \ref{fig:charge} and previous verification of the calibration \cite{Redford2018} we believe that there is a strong enough case to use the JUNGFRAU high gain as reference. The black dashed guides in the figure shows the expected variations due to the Poisson distribution of the incoming particles.

\section{Conclusions}

 With JUNGFRAU we can measure individual electrons with a high signal to noise ratio although the spatial resolution is somewhat limited by scattering in the sensor layer. Simulations show that moving to a thinner sensor would improve spatial resolution without reduced detection efficiency, due to the low noise of JUNGFRAU. A thinner sensor would also have the added benefit of giving higher dynamic range due to the lower energy deposition per primary electron. All in all, JUNGFRAU is an excellent detector for diffraction experiments at pulsed MeV electron sources and should deliver the same high quality data as it is currently doing at X-ray free electron lasers.  

% \clearpage
\section{Methods and material}

\subsection{The CLARA accelerator}
\label{sec:clara}
Located at the STFC Daresbury Lab the CLARA\cite{clara} accelerator offers up to 45 MeV beam with <5ps duration and a bunch charge in the range of 20 - 100 pC. Lower charges are possible, as shown in this paper, but without proper beam diagnostics. Currently the machine operates a 10 Hz but an upgrade program is underway increasing the energy, bunch charge and repetition rate for the next phase. 

\subsection{Clustering}
A cluster refers to the number of pixels triggered by a single incident electron. We performed the clustering using connected components labeling with 8-connectivity \cite{lifeng2017}, after first applying a 10 keV threshold to the pedestal and gain corrected image. Both the simulation and the measurement data was fed into the same processing pipeline. The drawback of this processing is that we risk overlapping cluster if the intensity is too high and in the rare occasion when the electron reenters the sensor after being scattered in the read-out ASIC we might count a single electron twice. 

\subsection{Measuring MTF}
\label{sec:mtf}
The MTF was measured using a slanted edge \cite{IEC62220} (3 mm tungsten). To minimise the effect of scattering on the results the edge was placed directly on the silicon sensor with only a thin Kapton tape in between for protection and electrical insulation. Due to the limited thickness we could only measure MTF at 4 and 10 MeV. To correct for the beam shape we applied a flat field correction before generating the super sampled edge. 

\subsection{Simulating the detector response}
\label{sec:sim}
\begin{figure}
    \centering
    \includegraphics[width=0.8\textwidth]{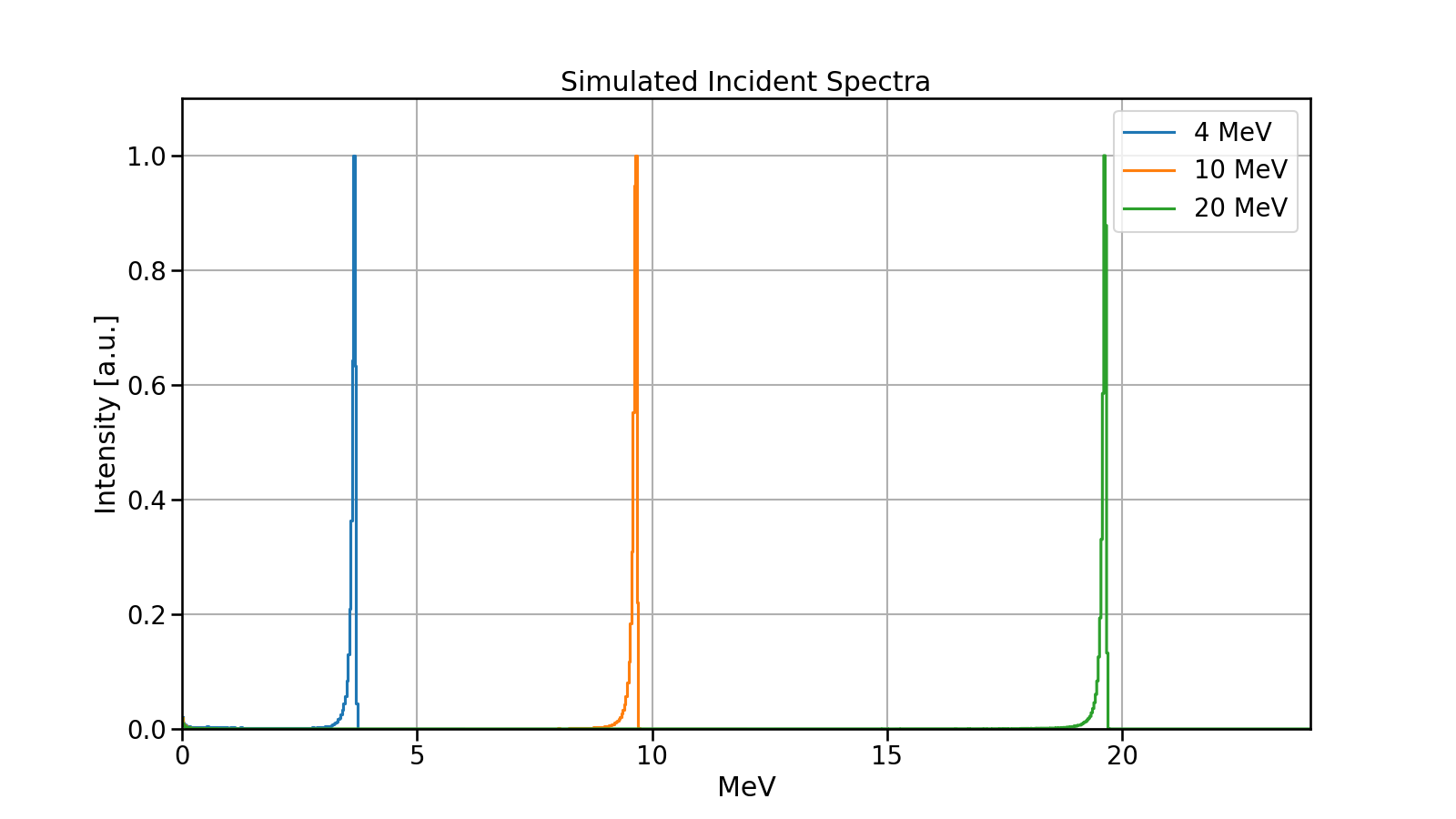}
    \caption{Simulated incident spectra. The beam energies used in this experiment were 4, 10 and 20 MeV. Energy loss due to the transport through the Be window and air was about 0.35 MeV.}
    \label{fig:incident}
\end{figure}

To better understand the detector performance and investigate further optimizations we simulated the measurement setup using a Geant4\cite{Agostinelli2003} based simulation framework that includes charge transport using drift-diffusion \cite{Schubel2014, Krapohl2016}. Since the beam has to pass through a 0.5 mm thick beryllium window in addition to 1.22 m of air (low intensity measurements) the simulations are also important to estimate the energy of the incident electrons and the effect scattering will have on the measurement results. Figure \ref{fig:incident} shows the simulated incident spectra for 4, 10 and 20 keV. During the transport the electrons loose about 0.35 MeV and the energy spread on the detector is 75 keV FWHM. 

\section*{Acknowledgements}
One of the authors (V. Hinger) has received funding from MSCA PSI-FELLOW-III-3i (EU grant agreement No. 884104)

\bibliographystyle{JHEP}
\bibliography{mybibfile}
\end{document}